\documentclass[sigconf]{acmart}
\AtBeginDocument{%
  \providecommand\BibTeX{{%
    \normalfont B\kern-0.5em{\scshape i\kern-0.25em b}\kern-0.8em\TeX}}}

\copyrightyear{2022}
\acmYear{2022}
\setcopyright{acmcopyright}
\acmConference[SIGIR '22]{Proceedings of the 45th
International ACM SIGIR Conference on Research and Development in Information
Retrieval}{July 11--15, 2022}{Madrid, Spain}
\acmBooktitle{Proceedings of the 45th International ACM SIGIR Conference on
Research and Development in Information Retrieval (SIGIR '22), July 11--15, 2022,
Madrid, Spain}
\acmPrice{15.00}
\acmDOI{10.1145/3477495.3531755}
\acmISBN{978-1-4503-8732-3/22/07}

\usepackage{booktabs} 
\usepackage{xspace}
\usepackage{subcaption}
\usepackage{tikz}
\usepackage{rotating}
\usepackage{pdflscape}
\usepackage{acronym}
\usepackage{multirow}

\newcommand{\idest}{i.e.,\xspace}
\newcommand{\eg}{e.g.,\xspace}

\newcommand{\tabref}[1]{Table \ref{#1}\xspace}

\acrodef{3G}[3G]{Third Generation Mobile System}
\acrodef{5S}[5S]{Streams, Structures, Spaces, Scenarios, Societies}
\acrodef{AA}[AA]{Active Agreements}
\acrodef{AAAI}[AAAI]{Association for the Advancement of Artificial Intelligence}
\acrodef{AAL}[AAL]{Annotation Abstraction Layer}
\acrodef{AAM}[AAM]{Automatic Annotation Manager}
\acrodef{AAP}[AAP]{Average Average Precision}
\acrodef{ACLIA}[ACLIA]{Advanced Cross-Lingual Information Access}
\acrodef{ACM}[ACM]{Association for Computing Machinery}
\acrodef{AD}[AD]{Active Disagreements}
\acrodef{ADSL}[ADSL]{Asymmetric Digital Subscriber Line}
\acrodef{ADUI}[ADUI]{ADministrator User Interface}
\acrodef{AIP}[AIP]{Archival Information Package}
\acrodef{AJAX}[AJAX]{Asynchronous JavaScript Technology and \acs{XML}}
\acrodef{ALU}[ALU]{Aritmetic-Logic Unit}
\acrodef{AMUSID}[AMUSID]{Adaptive MUSeological IDentity-service}
\acrodef{ANOVA}[ANOVA]{ANalysis Of VAriance}
\acrodef{ANSI}[ANSI]{American National Standards Institute}
\acrodef{AP}[AP]{Average Precision}
\acrodef{APC}[APC]{AP Correlation}
\acrodef{API}[API]{Application Program Interface}
\acrodef{AR}[AR]{Address Register}
\acrodef{AS}[AS]{Annotation Service}
\acrodef{ASAP}[ASAP]{Adaptable Software Architecture Performance}
\acrodef{ASI}[ASI]{Annotation Service Integrator}
\acrodef{ASL}[ASL]{Achieved Significance Level}
\acrodef{ASM}[ASM]{Annotation Storing Manager}
\acrodef{ASR}[ASR]{Automatic Speech Recognition}
\acrodef{ASUI}[ASUI]{ASsessor User Interface}
\acrodef{ATIM}[ATIM]{Annotation Textual Indexing Manager}
\acrodef{AUC}[AUC]{Area Under the ROC Curve}
\acrodef{AUI}[AUI]{Administrative User Interface}
\acrodef{AWARE}[AWARE]{Assessor-driven Weighted Averages for Retrieval Evaluation}
\acrodef{BANKS-I}[BANKS-I]{Browsing ANd Keyword Searching I}
\acrodef{BANKS-II}[BANKS-II]{Browsing ANd Keyword Searching II}
\acrodef{BH}[BH]{Benjamini-Hochberg}
\acrodef{bpref}[bpref]{Binary Preference}
\acrodef{BNF}[BNF]{Backus and Naur Form}
\acrodef{BRICKS}[BRICKS]{Building Resources for Integrated Cultural Knowledge Services}
\acrodef{CAN}[CAN]{Content Addressable Netword}
\acrodef{CAS}[CAS]{Content-And-Structure}
\acrodef{CBSD}[CBSD]{Component-Based Software Developlement}
\acrodef{CBSE}[CBSE]{Component-Based Software Engineering}
\acrodef{CB-SPE}[CB-SPE]{Component-Based \acs{SPE}}
\acrodef{CD}[CD]{Collaboration Diagram}
\acrodef{CD}[CD]{Compact Disk}
\acrodef{CDF}[CDF]{Cumulative Density Function}
\acrodef{CENL}[CENL]{Conference of European National Librarians}
\acrodef{CIDOC CRM}[CIDOC CRM]{CIDOC Conceptual Reference Model}
\acrodef{CIR}[CIR]{Current Instruction Register}
\acrodef{CIRCO}[CIRCO]{Coordinated Information Retrieval Components Orchestration}
\acrodef{CG}[CG]{Cumulated Gain}
\acrodef{CL}[CL]{Curriculum Learning}
\acrodef{CL-ESA}[CL-ESA]{Cross-Lingual Explicit Semantic Analysis}
\acrodef{CLAIRE}[CLAIRE]{Combinatorial visuaL Analytics system for Information Retrieval Evaluation}
\acrodef{CLEF1}[CLEF]{Cross-Language Evaluation Forum}
\acrodef{CLEF}[CLEF]{Conference and Labs of the Evaluation Forum}
\acrodef{CLIR}[CLIR]{Cross Language Information Retrieval}
\acrodef{CM}[CM]{Continuation Methods}
\acrodef{CMS}[CMS]{Content Management System}
\acrodef{CMT}[CMT]{Campaign Management Tool}
\acrodef{CNR}[CNR]{Italian National Council of Research}
\acrodef{CO}[CO]{Content-Only}
\acrodef{COD}[COD]{Code On Demand}
\acrodef{CODATA}[CODATA]{Committee on Data for Science and Technology}
\acrodef{COLLATE}[COLLATE]{Collaboratory for Annotation Indexing and Retrieval of Digitized Historical Archive Material}
\acrodef{CP}[CP]{Characteristic Pattern}
\acrodef{CPE}[CPE]{Control Processor Element}
\acrodef{CPU}[CPU]{Central Processing Unit}
\acrodef{CQL}[CQL]{Contextual Query Language}
\acrodef{CRP}[CRP]{Cumulated Relative Position}
\acrodef{CRUD}[CRUD]{Create--Read--Update--Delete}
\acrodef{CS}[CS]{Characteristic Structure}
\acrodef{CSM}[CSM]{Campaign Storing Manager}
\acrodef{CSS}[CSS]{Cascading Style Sheets}
\acrodef{CTR}[CTR]{Click-Through Rate}
\acrodef{CU}[CU]{Control Unit}
\acrodef{CUI}[CUI]{Client User Interface}
\acrodef{CV}[CV]{Cross-Validation}
\acrodef{DAFFODIL}[DAFFODIL]{Distributed Agents for User-Friendly Access of Digital Libraries}
\acrodef{DAO}[DAO]{Data Access Object}
\acrodef{DARE}[DARE]{Drawing Adequate REpresentations}
\acrodef{DARPA}[DARPA]{Defense Advanced Research Projects Agency}
\acrodef{DAS}[DAS]{Distributed Annotation System}
\acrodef{DB}[DB]{DataBase}
\acrodef{DBMS}[DBMS]{DataBase Management System}
\acrodef{DC}[DC]{Dublin Core}
\acrodef{DCG}[DCG]{Discounted Cumulated Gain}
\acrodef{DCMI}[DCMI]{Dublin Core Metadata Initiative}
\acrodef{DCV}[DCV]{Document Cut--off Value}
\acrodef{DD}[DD]{Deployment Diagram}
\acrodef{DDC}[DDC]{Dewey Decimal Classification}
\acrodef{DDS}[DDS]{Direct Data Structure}
\acrodef{DF}[DF]{Degrees of Freedom}
\acrodef{DFI}[DFI]{Divergence From Independence}
\acrodef{DFR}[DFR]{Divergence From Randomness}
\acrodef{DHT}[DHT]{Distributed Hash Table}
\acrodef{DI}[DI]{Digital Image}
\acrodef{DIKW}[DIKW]{Data, Information, Knowledge, Wisdom}
\acrodef{DIL}[DIL]{\acs{DIRECT} Integration Layer}
\acrodef{DiLAS}[DiLAS]{Digital Library Annotation Service}
\acrodef{DIRECT}[DIRECT]{Distributed Information Retrieval Evaluation Campaign Tool}
\acrodef{DKMS}[DKMS]{Data and Knowledge Management System}
\acrodef{DL}[DL]{Digital Library}
\acrodefplural{DL}[DL]{Digital Libraries}
\acrodef{DLMS}[DLMS]{Digital Library Management System}
\acrodef{DLOG}[DL]{Description Logics}
\acrodef{DLS}[DLS]{Digital Library System}
\acrodef{DLSS}[DLSS]{Digital Library Service System}
\acrodef{DM}[DM]{Data Mining}
\acrodef{DO}[DO]{Digital Object}
\acrodef{DOI}[DOI]{Digital Object Identifier}
\acrodef{DOM}[DOM]{Document Object Model}
\acrodef{DoMDL}[DoMDL]{Document Model for Digital Libraries}
\acrodef{DP}[DP]{Discriminative Power}
\acrodef{DPBF}[DPBF]{Dynamic Programming Best-First}
\acrodef{DR}[DR]{Data Register}
\acrodef{DRIVER}[DRIVER]{Digital Repository Infrastructure Vision for European Research}
\acrodef{DTD}[DTD]{Document Type Definition}
\acrodef{DVD}[DVD]{Digital Versatile Disk}
\acrodef{EAC-CPF}[EAC-CPF]{Encoded Archival Context for Corporate Bodies, Persons, and Families}
\acrodef{EAD}[EAD]{Encoded Archival Description}
\acrodef{EAN}[EAN]{International Article Number}
\acrodef{EBU}[EBU]{Expected Browsing Utility}
\acrodef{ECD}[ECD]{Enhanced Contenty Delivery}
\acrodef{ECDL}[ECDL]{European Conference on Research and Advanced Technology for Digital Libraries}
\acrodef{EDM}[EDM]{Europeana Data Model}
\acrodef{EG}[EG]{Execution Graph}
\acrodef{ELDA}[ELDA]{Evaluation and Language resources Distribution Agency}
\acrodef{ELRA}[ELRA]{European Language Resources Association}
\acrodef{EM}[EM]{Expectation Maximization}
\acrodef{EMMA}[EMMA]{Extensible MultiModal Annotation}
\acrodef{EPROM}[EPROM]{Erasable Programmable \acs{ROM}}
\acrodef{EQNM}[EQNM]{Extended Queueing Network Model}
\acrodef{ER}[ER]{Entity--Relationship}
\acrodef{ERR}[ERR]{Expected Reciprocal Rank}
\acrodef{ERS}[ERS]{Empirical Relational System}
\acrodef{ESA}[ESA]{Explicit Semantic Analysis}
\acrodef{ESL}[ESL]{Expected Search Length}
\acrodef{ETL}[ETL]{Extract-Transform-Load}
\acrodef{FAST}[FAST]{Flexible Annotation Service Tool}
\acrodef{FDR}[FDR]{False Discovery Rate}
\acrodef{FIFO}[FIFO]{First-In / First-Out}
\acrodef{FIRE}[FIRE]{Forum for Information Retrieval Evaluation}
\acrodef{FN}[FN]{False Negative}
\acrodef{FNR}[FNR]{False Negative Rate}
\acrodef{FOAF}[FOAF]{Friend of a Friend}
\acrodef{FORESEE}[FORESEE]{FOod REcommentation sErvER}
\acrodef{FP}[FP]{False Positive}
\acrodef{FPR}[FPR]{False Positive Rate}
\acrodef{FWER}[FWER]{Family-wise Error Rate}
\acrodef{GIF}[GIF]{Graphics Interchange Format}
\acrodef{GIR}[GIR]{Geografic Information Retrieval}
\acrodef{GAP}[GAP]{Graded Average Precision}
\acrodef{GLM}[GLM]{General Linear Model}
\acrodef{GLMM}[GLMM]{General Linear Mixed Model}
\acrodef{GMAP}[GMAP]{Geometric Mean Average Precision}
\acrodef{GoP}[GoP]{Grid of Points}
\acrodef{GPRS}[GPRS]{General Packet Radio Service}
\acrodef{gP}[gP]{Generalized Precision}
\acrodef{gR}[gR]{Generalized Recall}
\acrodef{gRBP}[gRBP]{Graded Rank-Biased Precision}
\acrodef{GT}[GT]{Generalizability Theory}
\acrodef{GTIN}[GTIN]{Global Trade Item Number}
\acrodef{GUI}[GUI]{Graphical User Interface}
\acrodef{GW}[GW]{Gateway}
\acrodef{HCI}[HCI]{Human Computer Interaction}
\acrodef{HDS}[HDS]{Hybrid Data Structure}
\acrodef{HIR}[HIR]{Hypertext Information Retrieval}
\acrodef{HIT}[HIT]{Human Intelligent Task}
\acrodef{HITS}[HITS]{Hyperlink-Induced Topic Search}
\acrodef{HMM}[HMM]{Hidden Markov Model}
\acrodef{HTML}[HTML]{HyperText Markup Language}
\acrodef{HTTP}[HTTP]{HyperText Transfer Protocol}
\acrodef{HSD}[HSD]{Honestly Significant Difference}
\acrodef{ICA}[ICA]{International Council on Archives}
\acrodef{ICSU}[ICSU]{International Council for Science}
\acrodef{IDF}[IDF]{Inverse Document Frequency}
\acrodef{IDS}[IDS]{Inverse Data Structure}
\acrodef{IEEE}[IEEE]{Institute of Electrical and Electronics Engineers}
\acrodef{IEI}[IEI]{Istituto della Enciclopedia Italiana fondata da Giovanni Treccani}
\acrodef{IETF}[IETF]{Internet Engineering Task Force}
\acrodef{IIR}[IIR]{Interactive Information Retrieval}
\acrodef{IMS}[IMS]{Information Management System}
\acrodef{IMSPD}[IMS]{Information Management Systems Research Group}
\acrodef{indAP}[indAP]{Induced Average Precision}
\acrodef{infAP}[infAP]{Inferred Average Precision}
\acrodef{INEX}[INEX]{INitiative for the Evaluation of \acs{XML} Retrieval}
\acrodef{INS-M}[INS-M]{Inverse Set Data Model}
\acrodef{INTR}[INTR]{Interrupt Register}
\acrodef{IP}[IP]{Internet Protocol}
\acrodef{IPSA}[IPSA]{Imaginum Patavinae Scientiae Archivum}
\acrodef{IR}[IR]{Information Retrieval}
\acrodef{IRON}[IRON]{Information Retrieval ON}
\acrodef{IRON2}[IRON$^2$]{Information Retrieval On aNNotations}
\acrodef{IRON-SAT}[IRON-SAT]{\acs{IRON} - Statistical Analysis Tool}
\acrodef{IRS}[IRS]{Information Retrieval System}
\acrodef{ISAD(G)}[ISAD(G)]{International Standard for Archival Description (General)}
\acrodef{ISBN}[ISBN]{International Standard Book Number}
\acrodef{ISIS}[ISIS]{Interactive SImilarity Search}
\acrodef{ISJ}[ISJ]{Interactive Searching and Judging}
\acrodef{ISO}[ISO]{International Organization for Standardization}
\acrodef{ITU}[ITU]{International Telecommunication Union }
\acrodef{ITU-T}[ITU-T]{Telecommunication Standardization Sector of \acs{ITU}}
\acrodef{IV}[IV]{Information Visualization}
\acrodef{JAN}[JAN]{Japanese Article Number}
\acrodef{JDBC}[JDBC]{Java DataBase Connectivity}
\acrodef{JMB}[JMB]{Java--Matlab Bridge}
\acrodef{JPEG}[JPEG]{Joint Photographic Experts Group}
\acrodef{JSON}[JSON]{JavaScript Object Notation}
\acrodef{JSP}[JSP]{Java Server Pages}
\acrodef{JTE}[JTE]{Java-Treceval Engine}
\acrodef{KDE}[KDE]{Kernel Density Estimation}
\acrodef{KLD}[KLD]{Kullback-Leibler Divergence}
\acrodef{KLAPER}[KLAPER]{Kernel LAnguage for PErformance and Reliability analysis}
\acrodef{LAM}[LAM]{Libraries, Archives, and Museums}
\acrodef{LAM2}[LAM]{Logistic Average Misclassification}
\acrodef{LAN}[LAN]{Local Area Network}
\acrodef{LD}[LD]{Linked Data}
\acrodef{LEAF}[LEAF]{Linking and Exploring Authority Files}
\acrodef{LIDO}[LIDO]{Lightweight Information Describing Objects}
\acrodef{LIFO}[LIFO]{Last-In / First-Out}
\acrodef{LM}[LM]{Language Model}
\acrodef{LMT}[LMT]{Log Management Tool}
\acrodef{LOD}[LOD]{Linked Open Data}
\acrodef{LODE}[LODE]{Linking Open Descriptions of Events}
\acrodef{LpO}[LpO]{Leave-$p$-Out}
\acrodef{LRM}[LRM]{Local Relational Model}
\acrodef{LRU}[LRU]{Last Recently Used}
\acrodef{LS}[LS]{Lexical Signature}
\acrodef{LSM}[LSM]{Log Storing Manager}
\acrodef{LtR}[LtR]{Learning to Rank}
\acrodef{LUG}[LUG]{Lexical Unit Generator}
\acrodef{MA}[MA]{Mobile Agent}
\acrodef{MA}[MA]{Moving Average}
\acrodef{MACS}[MACS]{Multilingual ACcess to Subjects}
\acrodef{MADCOW}[MADCOW]{Multimedia Annotation of Digital Content Over the Web}
\acrodef{MAD}[MAD]{Mean Assessed Documents}
\acrodef{MADP}[MADP]{Mean Assessed Documents Precision}
\acrodef{MADS}[MADS]{Metadata Authority Description Standard}
\acrodef{MAP}[MAP]{Mean Average Precision}
\acrodef{MARC}[MARC]{Machine Readable Cataloging}
\acrodef{MATTERS}[MATTERS]{MATlab Toolkit for Evaluation of information Retrieval Systems}
\acrodef{MDA}[MDA]{Model Driven Architecture}
\acrodef{MDD}[MDD]{Model-Driven Development}
\acrodef{METS}[METS]{Metadata Encoding and Transmission Standard}
\acrodef{MIDI}[MIDI]{Musical Instrument Digital Interface}
\acrodef{MIME}[MIME]{Multipurpose Internet Mail Extensions}
\acrodef{MIQUBO}[MIQUBO]{Mutual Information QUBO}
\acrodef{ML}[ML]{Machine Learning}
\acrodef{MLE}[MLE]{Maximum Likelihood Estimation}
\acrodef{MLIA}[MLIA]{MultiLingual Information Access}
\acrodef{MM}[MM]{Machinery Model}
\acrodef{MMU}[MMU]{Memory Management Unit}
\acrodef{MODS}[MODS]{Metadata Object Description Schema}
\acrodef{MOF}[MOF]{Meta-Object Facility}
\acrodef{MP}[MP]{Markov Precision}
\acrodef{MPEG}[MPEG]{Motion Picture Experts Group}
\acrodef{MRD}[MRD]{Machine Readable Dictionary}
\acrodef{MRF}[MRF]{Markov Random Field}
\acrodef{MRR}[MRR]{Mean Reciprocal Rank}
\acrodef{MS}[MS]{Mean Squares}
\acrodef{MSAC}[MSAC]{Multilingual Subject Access to Catalogues}
\acrodef{MSE}[MSE]{Mean Square Error}
\acrodef{MT}[MT]{Machine Translation}
\acrodef{MV}[MV]{Majority Vote}
\acrodef{MVC}[MVC]{Model-View-Controller}
\acrodef{NACSIS}[NACSIS]{NAtional Center for Science Information Systems}
\acrodef{NAP}[NAP]{Network processors Applications Profile}
\acrodef{NCP}[NCP]{Normalized Cumulative Precision}
\acrodef{nCG}[nCG]{Normalized Cumulated Gain}
\acrodef{nCRP}[nCRP]{Normalized Cumulated Relative Position}
\acrodef{nDCG}[nDCG]{Normalized Discounted Cumulated Gain}
\acrodef{nMCG}[nMCG]{Normalized Markov Cumulated Gain}
\acrodef{NESTOR}[NESTOR]{NEsted SeTs for Object hieRarchies}
\acrodef{NEXI}[NEXI]{Narrowed Extended XPath I}
\acrodef{NII}[NII]{National Institute of Informatics}
\acrodef{NISO}[NISO]{National Information Standards Organization}
\acrodef{NIST}[NIST]{National Institute of Standards and Technology}
\acrodef{NLP}[NLP]{Natural Language Processing}
\acrodef{NN}[NN]{Neural Network}
\acrodef{NP}[NP]{Network Processor}
\acrodef{NR}[NR]{Normalized Recall}
\acrodef{NRS}[NRS]{Numerical Relational System}
\acrodef{NS-M}[NS-M]{Nested Set Model}
\acrodef{NTCIR}[NTCIR]{NII Testbeds and Community for Information access Research}
\acrodef{OAI}[OAI]{Open Archives Initiative}
\acrodef{OAI-ORE}[OAI-ORE]{Open Archives Initiative Object Reuse and Exchange}
\acrodef{OAI-PMH}[OAI-PMH]{Open Archives Initiative Protocol for Metadata Harvesting}
\acrodef{OAIS}[OAIS]{Open Archival Information System}
\acrodef{OC}[OC]{Operation Code}
\acrodef{OCLC}[OCLC]{Online Computer Library Center}
\acrodef{OMG}[OMG]{Object Management Group}
\acrodef{OO}[OO]{Object Oriented}
\acrodef{OODB}[OODB]{Object-Oriented \acs{DB}}
\acrodef{OODBMS}[OODBMS]{Object-Oriented \acs{DBMS}}
\acrodef{OPAC}[OPAC]{Online Public Access Catalog}
\acrodef{OQL}[OQL]{Object Query Language}
\acrodef{ORP}[ORP]{Open Relevance Project}
\acrodef{OSIRIS}[OSIRIS]{Open Service Infrastructure for Reliable and Integrated process Support}
\acrodef{P}[P]{Precision}
\acrodef{P2P}[P2P]{Peer-To-Peer}
\acrodef{PA}[PA]{Passive Agreements}
\acrodef{PAMT}[PAMT]{Pool-Assessment Management Tool}
\acrodef{PASM}[PASM]{Pool-Assessment Storing Manager}
\acrodef{PC}[PC]{Program Counter}
\acrodef{PCP}[PCP]{Pre-Commercial Procurement}
\acrodef{PCR}[PCR]{Peripherical Command Register}
\acrodef{PD}[PD]{Passive Disagreements}
\acrodef{PDA}[PDA]{Personal Digital Assistant}
\acrodef{PDF}[PDF]{Probability Density Function}
\acrodef{PDR}[PDR]{Peripherical Data Register}
\acrodef{PIR}[PIR]{Personalized Information Retrieval}
\acrodef{POI}[POI]{\acs{PURL}-based Object Identifier}
\acrodef{PoS}[PoS]{Part of Speech}
\acrodef{PAA}[PAA]{Proportion of Active Agreements}
\acrodef{PPA}[PPA]{Proportion of Passive Agreements}
\acrodef{PPE}[PPE]{Programmable Processing Engine}
\acrodef{PREFORMA}[PREFORMA]{PREservation FORMAts for culture information/e-archives}
\acrodef{PRIMAD}[PRIMAD]{Platform, Research goal, Implementation, Method, Actor, and Data}
\acrodef{PRIMAmob-UML}[PRIMAmob-UML]{mobile \acs{PRIMA-UML}}
\acrodef{PRIMA-UML}[PRIMA-UML]{PeRformance IncreMental vAlidation in \acs{UML}}
\acrodef{PROM}[PROM]{Programmable \acs{ROM}}
\acrodef{PROMISE}[PROMISE]{Participative Research labOratory  for Multimedia and Multilingual Information Systems Evaluation}
\acrodef{pSQL}[pSQL]{propagate \acs{SQL}}
\acrodef{PUI}[PUI]{Participant User Interface}
\acrodef{PURL}[PURL]{Persistent \acs{URL}}
\acrodef{QA}[QA]{Quantum Annealing}
\acrodef{QE}[QE]{Query Expansion}
\acrodef{QoS-UML}[QoS-UML]{\acs{UML} Profile for QoS and Fault Tolerance}
\acrodef{QPA}[QPA]{Query Performance Analyzer}
\acrodef{QPP}[QPP]{Query Performance Prediction}
\acrodef{QPU}[QPU]{Quantum Processing Unit}
\acrodef{QUBO}[QUBO]{Quadratic Unconstrained Binary Optimization}
\acrodef{R}[R]{Recall}
\acrodef{RAM}[RAM]{Random Access Memory}
\acrodef{RAMM}[RAM]{Random Access Machine}
\acrodef{RBO}[RBO]{Rank-Biased Overlap}
\acrodef{RBP}[RBP]{Rank-Biased Precision}
\acrodef{RBTO}[RBTO]{Rank-Based Total Order}
\acrodef{RDBMS}[RDBMS]{Relational \acs{DBMS}}
\acrodef{RDF}[RDF]{Resource Description Framework}
\acrodef{REST}[REST]{REpresentational State Transfer}
\acrodef{REV}[REV]{Remote Evaluation}
\acrodef{RF}[RF]{Relevance Feedback}
\acrodef{RFC}[RFC]{Request for Comments}
\acrodef{RIA}[RIA]{Reliable Information Access}
\acrodef{RMSE}[RMSE]{Root Mean Square Error}
\acrodef{RMT}[RMT]{Run Management Tool}
\acrodef{ROM}[ROM]{Read Only Memory}
\acrodef{ROMIP}[ROMIP]{Russian Information Retrieval Evaluation Seminar}
\acrodef{RoMP}[RoMP]{Rankings of Measure Pairs}
\acrodef{RoS}[RoS]{Rankings of Systems}
\acrodef{RP}[RP]{Relative Position}
\acrodef{RR}[RR]{Reciprocal Rank}
\acrodef{RSM}[RSM]{Run Storing Manager}
\acrodef{RST}[RST]{Rhetorical Structure Theory}
\acrodef{RSV}[RSV]{Retrieval Status Value}
\acrodef{RT-UML}[RT-UML]{\acs{UML} Profile for Schedulability, Performance and Time}
\acrodef{SA}[SA]{Simulated Annealing}
\acrodef{SAL}[SAL]{Storing Abstraction Layer}
\acrodef{SAMT}[SAMT]{Statistical Analysis Management Tool}
\acrodef{SAN}[SAN]{Sistema Archivistico Nazionale}
\acrodef{SASM}[SASM]{Statistical Analysis Storing Manager}
\acrodef{SBTO}[SBTO]{Set-Based Total Order}
\acrodef{SD}[SD]{Steepest Descent}
\acrodef{SE}[SE]{Search Engine}
\acrodef{SEBD}[SEBD]{Convegno Nazionale su Sistemi Evoluti per Basi di Dati}
\acrodef{SEM}[SEM]{Standard Error of the Mean}
\acrodef{SERP}[SERP]{Search Engine Result Page}
\acrodef{SFT}[SFT]{Satisfaction--Frustration--Total}
\acrodef{SIL}[SIL]{Service Integration Layer}
\acrodef{SIP}[SIP]{Submission Information Package}
\acrodef{SKOS}[SKOS]{Simple Knowledge Organization System}
\acrodef{SM}[SM]{Software Model}
\acrodef{SME}[SME]{Statistics--Metrics-Experiments}
\acrodef{SMART}[SMART]{System for the Mechanical Analysis and Retrieval of Text}
\acrodef{SoA}[SoA]{Service-oriented Architectures}
\acrodef{SOA}[SOA]{Strength of Association}
\acrodef{SOAP}[SOAP]{Simple Object Access Protocol}
\acrodef{SOM}[SOM]{Self-Organizing Map}
\acrodef{SPARQL}[SPARQL]{Simple Protocol and RDF Query Language}
\acrodef{SPE}[SPE]{Software Performance Engineering}
\acrodef{SPINA}[SPINA]{Superimposed Peer Infrastructure for iNformation Access}
\acrodef{SPLIT}[SPLIT]{Stemming Program for Language Independent Tasks}
\acrodef{SPOOL}[SPOOL]{Simultaneous Peripheral Operations On Line}
\acrodef{SQL}[SQL]{Structured Query Language}
\acrodef{SR}[SR]{Sliding Ratio}
\acrodef{sRBP}[sRBP]{Session Rank Biased Precision}
\acrodef{SRU}[SRU]{Search/Retrieve via \acs{URL}}
\acrodef{SS}[SS]{Sum of Squares}
\acrodef{SSD}[s.s.d.]{statistically significantly different}
\acrodef{SSTF}[SSTF]{Shortest Seek Time First}
\acrodef{STAR}[STAR]{Steiner-Tree Approximation in Relationship graphs}
\acrodef{STON}[STON]{STemming ON}
\acrodef{SVC}[SVC]{Support Vector Classifier}
\acrodef{SVM}[SVM]{Support Vector Machine}
\acrodef{TAC}[TAC]{Text Analysis Conference}
\acrodef{TBG}[TBG]{Time-Biased Gain}
\acrodef{TCP}[TCP]{Transmission Control Protocol}
\acrodef{TEL}[TEL]{The European Library}
\acrodef{TERRIER}[TERRIER]{TERabyte RetrIEveR}
\acrodef{TF}[TF]{Term Frequency}
\acrodef{TFR}[TFR]{True False Rate}
\acrodef{TLD}[TLD]{Top Level Domain}
\acrodef{TME}[TME]{Topics--Metrics-Experiments}
\acrodef{TN}[TN]{True Negative}
\acrodef{TO}[TO]{Transfer Object}
\acrodef{TP}[TP]{True Positve}
\acrodef{TPR}[TPR]{True Positive Rate}
\acrodef{TRAT}[TRAT]{Text Relevance Assessing Task}
\acrodef{TREC}[TREC]{Text REtrieval Conference}
\acrodef{TRECVID}[TRECVID]{TREC Video Retrieval Evaluation}
\acrodef{TS}[TS]{Tabu Search}
\acrodef{TTL}[TTL]{Time-To-Live}
\acrodef{UCD}[UCD]{Use Case Diagram}
\acrodef{UDC}[UDC]{Universal Decimal Classification}
\acrodef{uGAP}[uGAP]{User-oriented Graded Average Precision}
\acrodef{UI}[UI]{User Interface}
\acrodef{UML}[UML]{Unified Modeling Language}
\acrodef{UMT}[UMT]{User Management Tool}
\acrodef{UMTS}[UMTS]{Universal Mobile Telecommunication System}
\acrodef{UoM}[UoM]{Utility-oriented Measurement}
\acrodef{UPC}[UPC]{Universal Product Code}
\acrodef{URI}[URI]{Uniform Resource Identifier}
\acrodef{URL}[URL]{Uniform Resource Locator}
\acrodef{URN}[URN]{Uniform Resource Name}
\acrodef{USM}[USM]{User Storing Manager}
\acrodef{VA}[VA]{Visual Analytics}
\acrodef{VAIRE}[VAIR\"{E}]{Visual Analytics for Information Retrieval Evaluation}
\acrodef{VATE}[VATE$^2$]{Visual Analytics Tool for Experimental Evaluation}
\acrodef{VIRTUE}[VIRTUE]{Visual Information Retrieval Tool for Upfront Evaluation}
\acrodef{VD}[VD]{Virtual Document}
\acrodef{VDM}[VDM]{Visual Data Mining}
\acrodef{VIAF}[VIAF]{Virtual International Authority File}
\acrodef{VIM}[VIM]{International Vocabulary of Metrology}
\acrodef{VL}[VL]{Visual Language}
\acrodef{VoIP}[VoIP]{Voice over IP}
\acrodef{VS}[VS]{Visual Sentence}
\acrodef{W3C}[W3C]{World Wide Web Consortium}
\acrodef{WAN}[WAN]{Wide Area Network}
\acrodef{WHO}[WHO]{World Health Organization}
\acrodef{WLAN}[WLAN]{Wireless \acs{LAN}}
\acrodef{WP}[WP]{Work Package}
\acrodef{WS}[WS]{Web Services}
\acrodef{WSD}[WSD]{Word Sense Disambiguation}
\acrodef{WSDL}[WSDL]{Web Services Description Language}
\acrodef{WWW}[WWW]{World Wide Web}
\acrodef{XMI}[XMI]{\acs{XML} Metadata Interchange}
\acrodef{XML}[XML]{eXtensible Markup Language}
\acrodef{XPath}[XPath]{XML Path Language}
\acrodef{XSL}[XSL]{eXtensible Stylesheet Language}
\acrodef{XSL-FO}[XSL-FO]{\acs{XSL} Formatting Objects}
\acrodef{XSLT}[XSLT]{\acs{XSL} Transformations}
\acrodef{YAGO}[YAGO]{Yet Another Great Ontology}
\acrodef{YASS}[YASS]{Yet Another Suffix Stripper}
\acrodef{KPI}[KPI]{Key Performance Indicator}
\acrodef{KNRM}[KNRM]{Kernel-based Neural Reranking Model}
\acrodef{DRMM}[DRMM]{Deep relevance Matching Model}
\acrodef{PACRR}[PACRR]{Position-Aware Convolutional Recurrent Relevance matching}
\acrodef{RM}[RM]{Relevance Model}
\acrodef{MAE}[MAE]{Mean Absolute Error}
\acrodef{CI}[CI]{Confidence Interval}

\begin{document}
\fancyhead{}


\title{Towards Feature Selection for Ranking and Classification Exploiting Quantum Annealers}

\author{Maurizio {Ferrari Dacrema}}
\orcid{0000-0001-7103-2788}
\affiliation{%
  \institution{Politecnico di Milano}
  \country{Italy}
}
\email{maurizio.ferrari@polimi.it}

\author{Fabio Moroni}
\affiliation{%
  \institution{Politecnico di Milano}
  \country{Italy}
}
\email{fabio6.moroni@mail.polimi.it}

\author{Riccardo Nembrini}
\orcid{0000-0002-1915-6107}
\affiliation{%
  \institution{Politecnico di Milano, ContentWise}
  \country{Italy}
}
\email{riccardo.nembrini@polimi.it}

\author{Nicola Ferro}
\orcid{0000-0001-9219-6239}
\affiliation{%
  \institution{Università degli Studi di Padova}
  \country{Italy}
}
\email{ferro@dei.unipd.it}

\author{Guglielmo Faggioli}
\orcid{0000-0002-5070-2049}
\affiliation{%
  \institution{Università degli Studi di Padova}
  \country{Italy}
}
\email{faggioli@dei.unipd.it}

\author{Paolo Cremonesi}
\orcid{0000-0002-1253-8081}
\affiliation{%
  \institution{Politecnico di Milano}
  \country{Italy}
}
\email{paolo.cremonesi@polimi.it}


\begin{abstract}
Feature selection is a common step in many ranking, classification, or prediction tasks and serves many purposes. By removing redundant or noisy features, the accuracy of ranking or classification can be improved and the computational cost of the subsequent learning steps can be reduced. However, feature selection can be itself a computationally expensive process. While for decades confined to theoretical algorithmic papers, quantum computing is now becoming a viable tool to tackle realistic problems, in particular special-purpose solvers based on the Quantum Annealing paradigm. This paper aims to explore the feasibility of using currently available quantum computing architectures to solve some quadratic feature selection algorithms for both ranking and classification.

The experimental analysis includes 15 state-of-the-art datasets. The effectiveness obtained with quantum computing hardware is comparable to that of classical solvers, indicating that quantum computers are now reliable enough to tackle interesting problems. In terms of scalability, current generation quantum computers are able to provide a limited speedup over certain classical algorithms and hybrid quantum-classical strategies show lower computational cost for problems of more than a thousand features.
 
\end{abstract}

\ccsdesc[500]{Information systems~Content analysis and feature selection}
\ccsdesc[500]{Computer systems organization~Quantum computing}

%
\keywords{Feature Selection, Quantum Computing, Quantum Annealing, Machine Learning}

\maketitle

\section{Introduction}

\acf{IR} is concerned with delivering relevant information to people, according to their information needs, context, and profile, in the most effective and efficient way possible. Central to this goal are \emph{ranking} and \emph{classification}, often exploited in conjunction. 
Machine learning approaches have been widely investigated for this purpose. These methods however suffer from the known feature selection problem. As the data becomes more rich and complex, identifying the relevant features may require to evaluate an exponentially increasing number of cases which rapidly becomes prohibitively resource intensive.  
The feature selection problem is mitigated by deep learning and, more generally, neural approaches that have gained popularity in recent years. Despite these methods being extremely versatile and generally able to provide good overall effectiveness, it is known their performance is not always stable and may vary a lot across topics, for example the performance may improve for half of the topics while degrade for the other half ~\cite{MarchesinEtAl2020}.
A further disadvantage is that these neural approaches are very demanding in terms of computing resources and require enormous amounts of data which leads to larger and larger models that are not free from risks, as pointed out by~\citet{BenderEtAl2021}.

In this paper we take a step back and wonder ourselves if it is possible to make the feature selection problem more ``affordable'' in order to make more appealing the use of ``traditional'' machine learning approaches for ranking and classification. To this end, we investigate the feasibility of and how to apply current generation quantum computing technologies to improve feature selection. To the best of our knowledge very little work has been done to asses the effectiveness and efficiency of such technologies to tackle feature selection problems, especially for both ranking and classification. For example, \citet{DBLP:journals/entropy/NembriniDC21} proposed a heuristic to build a hybrid recommender system that selects important features according to how well they allow us to approximate another recommendation model based on the user behavior. 

The contributions of this paper are the following:
\begin{itemize}
    \item formulate the feature selection problem as a \acf{QUBO} problem which can be solved using \acf{QA};
    \item compare efficiency and effectiveness of classical and quantum \ac{QUBO} problems solvers;
    \item show that a quantum computer is able to more efficiently solve the feature selection problem, for both ranking and classification, with an effectiveness comparable to classical solvers;
    \item show that the quantum-classical hybrid approaches exhibit better scalability when the number of features increases compared to the classical ones.
\end{itemize}

The results reported in this paper show that quantum computing approaches have become a viable option for the feature selection problem in \ac{IR} and that they are worthy of further investigation.

The paper is organized as follows: Section \ref{sec:related} describes related work on quantum computing and feature selection, Section \ref{sec:methodology} presents feature selection as a Quadratic Unconstrained Binary Optimization problem as well as classical and quantum strategies to solve such problems, Section \ref{sec:exprimental-pipeline} describes the experimental pipeline, Section \ref{sec:results} presents the results, finally Section \ref{sec:discussion} draws the conclusions and presents future research directions.

\section{Related Works}
\label{sec:related}

\paragraph{Quantum Computing}

In recent years, quantum computing is gaining popularity as a new computational paradigm able to offer speedups for several computational tasks that are difficult on classical hardware. Quantum computing is often used to refer to a computational paradigm called \emph{gate model} or \emph{universal quantum computer}, in which a quantum state is manipulated with a controlled sequence of operations performed via quantum gates, in a way that is not dissimilar from how classical computers operate. Universal quantum computers are theoretically able to compute any function and have rigorously proven speedups for certain tasks. On the other hand, they are currently quite unreliable due to noise and are available only with a limited number of qubits, the basic unit of quantum information. Furthermore, this paradigm requires specially designed algorithms that is challenging to develop. 

Another paradigm of quantum computing is
\acf{QA} in which a special purpose device, called Quantum Annealer, is used to rapidly sample optimal solutions of a \ac{QUBO} problem; note that the detailed description of \ac{QUBO} methods is deferred to subsection \ref{sec:QUBO_feature_selection}. As opposed to universal quantum computers, Quantum Annealers have a limited computational flexibility, being able only to tackle optimization problems, and the question of proving the type of speedup they offer is still open. On the other hand, they suffer much less from noise and are available with a rather large number of qubits, allowing to tackle problems of interesting size. Moreover, the simplicity of the required \ac{QUBO} formulation and fast solution time have made it a promising and widely accessible technology.

This has fueled significant research from both industry and academia to explore the potential of this technology. Several \ac{QUBO} formulations for important problems have been developed such as graph partitioning \cite{DBLP:journals/corr/Ushijima-Mwesigwa17,DBLP:conf/lwa/BauckhagePSHW19}, Support Vector Machines \cite{DBLP:journals/cphysics/WillschWRM20}, Restricted Boltzmann Machines \cite{DBLP:journals/corr/AdachiH15,amin2018quantum} and optimization for resource allocation \cite{Carugno2022}. \ac{QA} has also been applied to collaborative filtering \cite{EQTC_collaborative} and the personalization of the user interface in a recommender system \cite{DBLP:conf/recsys/DacremaFC21}.

For these reasons, this paper investigates feature selection via \ac{QA} and \ac{QUBO} formulation.

\paragraph{Feature Selection}

Feature selection is one of the problems that fit well with the mathematical formulation required to use a Quantum Annealer. Its goal is to isolate a subset of all available features in order to improve the effectiveness of a model, being it for ranking or classification, and/or to reduce the computational cost. Feature selection is a widely researched topic \cite{DBLP:journals/jmlr/Rodriguez-LujanHEC10,DBLP:conf/mipro/JovicBB15, DBLP:journals/cee/ChandrashekarS14} and the main approaches can be divided in three categories:
\begin{description}
    \item[Filter methods:] the features are selected based on information-theoretical measures that do not optimize the model itself, \eg variance or entropy.
    \item[Embedded methods:] the selection, or weighting, of the features is a part of the model training, \eg lasso or ridge regression, neural networks and factorization machines all learn a weight for the input feature data. 
    \item[Wrapper methods:] the model is considered as a black box and the features are selected as those that optimize the end effectiveness of the model.
\end{description}

This work focuses on filter methods, since they are agnostic about the model to optimize and are well suited to study and compare several \ac{QA} and \ac{QUBO} methods across a variety of models.

Filter methods for features selection can be further categorized into linear and quadratic approaches. The former consider individual features singularly while the latter also account for feature interaction. Among the most common linear approaches, we can list the following:
\begin{description}
    \item[ANOVA F-Test:] the Analysis of Variance \cite{DBLP:conf/ideal/Sanchez-MaronoCCA06, DHANYA20201561, DBLP:journals/tifs/BihlBT16} method attempts to minimize false negative errors.
    
    \item[Chi2 Test:] features are ranked according to their Chi-square statistic value. This test aims to identify the features that are more likely independent from the target label.
    
    \item[Mutual Information:] features are ranked according to their Mutual Information with respect to the target variable. The Mutual Information is a statistical measure also known as \textit{Shannon's Information} \cite{DBLP:journals/pr/ZengZZY15, DBLP:journals/pami/PengLD05, DBLP:journals/jmlr/Torkkola03}.
    
    \item[Pearson Correlation:] features are ranked according to their \textit{Pearson Correlation} with the target variable. 
    
    \item[Linear Boosting:] each feature is used to train a Support Vector Classifier, which is then used to label the samples. Features are ranked according to the \textit{Pearson Correlation} between the prediction of the classifier and the target variable.
    
    \item[Variance Threshold:] removes low-variance features, the feature score is computed as the variance of its values.

\end{description}

These approaches will be considered as plain baselines in the subsequent experiments. In Section~\ref{sec:QUBO_feature_selection} we will describe quadratic approaches in detail, since they are the focus of our \ac{QUBO} optimization.

\paragraph{Quantum Information Retrieval} Quantum \ac{IR} is a branch of \ac{IR} initiated by~\citet{vanRijsbergen2004} which exploits the formalism and concepts of quantum mechanics, such as Hilbert spaces and Hermitian operators, to formulate and describe \ac{IR} models and specific problems, such has relevance feedback. Over the years Quantum \ac{IR} has evolved~\cite{Melucci2015}, exploring mainly two areas -- (i) representation and ranking; (ii) user interaction -- leveraging ``Hilbert space models for representation learning and quantum probability rules to model cognitive interference in document ranking''~\cite{UpretyEtAl2021}.

This work does not deal with Quantum \ac{IR} but rather explores how Quantum Computing, in particular \acl{QA}, can be applied to directly solve the feature selection problem in ranking and classification for \ac{IR}.

\paragraph{Efficient \ac{LtR}} 
A further line of work strongly connected to this work concerns the study of efficient approaches for feature selection applied to the \ac{LtR} task. Among the most prominent efforts in this field, we list the one from \citet{GigliEtAl2016}. In \cite{GigliEtAl2016}, the authors propose to use the Hierarchical agglomerative Clustering Algorithm for feature Selection (HCAS) to select features to be fed to a \ac{LtR} algorithm in both a fast and effective way. 
Following a similar line of thoughts, \citet{LuccheseNardini2017} highlight the importance of feature selection to obtain computationally efficient \ac{LtR} solutions. In particular, in \cite{LuccheseNardini2017}, the authors reiterate the extent of the trade-off between efficiency and effectiveness that characterize \ac{LtR} and the role of features selection in improving the former without excessively degrading the latter. Finally, \citet{lai2013fsmrank} propose an optimization strategy for feature selection while \citet{purpura2021neural} exploit a neural networks based approach. Both endeavours mentioned above do not focus on efficiency aspects but highlight the improved effectiveness of \ac{LtR} approaches if used in combination with feature selection solutions.

\section{Methodology}
\label{sec:methodology}

\subsection{Feature Selection as a Quadratic Problem}
\label{sec:QUBO_feature_selection}

Identifying the appropriate features is not trivial and depends on the data, task and context of interest, \eg obtain the best possible model accuracy, discard correlated features to reduce redundancy, minimize the number of selected features to improve scalability, and so on.

Quadratic models for feature weighting and selection have been studied for several years~\cite{DBLP:journals/jmlr/Rodriguez-LujanHEC10,DBLP:journals/eswa/KatrutsaS17}. While \emph{feature weighting} is a simpler task, since when the problem coefficients are semi-definite positive the optimization problem is convex and can be solved rapidly, \emph{feature selection} is an NP-hard problem. In this work, we focus only on feature selection, since it is the hardest task and can possibly lead to bigger gains. This section describes three quadratic models for feature selection represented with the \ac{QUBO} formulation of optimization problems, which allows us to easily describe many important NP-complete and NP-hard problems \cite{10.3389/fphy.2014.00005, DBLP:journals/4or/GloverKD19}.
The \ac{QUBO} formulation is defined as follows:
\begin{equation*}
	\label{eq:QUBO}
	\begin{aligned}
		\min \quad & y = x^TQx
	\end{aligned}
\end{equation*}
where $ x \in \{0, 1\}^m $ is a vector of $ m $ binary variables and $ Q $ is an $m \times m$ matrix that defines the function to optimize. In all the following methods, each feature will be associated to a binary variable to indicate whether it should be selected or not, therefore $m=|F|$ with $F$ the set of existing features.

\subsubsection{\ac{MIQUBO}}
\label{sec:MIQUBO}
\ac{MIQUBO} is a quadratic feature selection model based on Mutual Information. Its objective is to maximize the mutual information between selected features and the target variable as well as the conditional mutual information between a feature and the target, given other selected features.
\ac{MIQUBO} is defined as:
\begin{equation*}
    Q_{ij} = 
    \begin{cases}
        -MI(f_i, y | f_j) & if \:\: i \neq j \\
        -MI(f_i, y) & if \:\: i = j
    \end{cases} 
    \label{eqn:MIQUBO}
\end{equation*}
where $MI(f_i, y)$ is the mutual information~\cite{DBLP:journals/pr/ZengZZY15, DBLP:journals/pami/PengLD05, DBLP:journals/jmlr/Torkkola03} between feature $f_i$ and target $y$ and  $MI(f_i, y | f_j)$ is the conditional mutual information between feature $f_i$ and target $y$ given feature $f_j$. 
Since mutual information is always non-negative, \ac{MIQUBO} has a trivial solution where all features are selected. Due to this \ac{MIQUBO} requires to specify what is the desired number of feature to select by introducing a penalization term.

\subsubsection{\ac{QUBO}-Correlation}
\label{sec:QUBO-Correlation}
This strategy, originally proposed by \citet{DBLP:journals/prl/FerreiraF12}, is based on the Pearson Correlation. \ac{QUBO}-Correlation aims to maximize the correlation between the selected features and the target variable, but also to minimize the correlation among the selected features in order to reduce redundancy. \ac{QUBO}-Correlation is defined as:
\begin{equation*}
    Q_{ij} = 
    \begin{cases}
        - r(f_i, f_j) &  if \:\: i \neq j \\
        r(f_i, y) & if \:\: i = j
    \end{cases} 
    \label{eqn:QUBO-Correlation}
\end{equation*}
where $r(\cdot, \cdot)$ is the Pearson's $r$ Correlation.

\subsubsection{\ac{QUBO}-Boosting}
\label{sec:QUBO-Boosting}
This technique estimates the information content of a feature based on the predictions computed by a classifier using only that feature and is inspired by the work done by \citet{DBLP:journals/jmlr/NevenDRM12}. 
\ac{QUBO}-Boosting operates by first training $|F|$ different \ac{SVC}, one per each existing feature. 
The trained classifier is then used to compute the predicted class for each sample. The correlation between the predicted label and the correct one is used to assess the information content of the feature. 
\ac{QUBO}-Boosting is defined as follows:
\begin{equation*}
    Q_{ij} = 
    \begin{cases}
        r(h_i, h_j) &  if \:\: i \neq j \\
        \frac{S}{|F|^2} + \lambda - 2*r(h_i, y) & if \:\: i = j
    \end{cases} 
    \label{eqn:QUBO-Boosting}
\end{equation*}
where $h_i$ is the predicted classification of the \ac{SVC} trained only on feature $i$, $S$ is the number of samples in the dataset and $\lambda$ is a hyperparameter. Note that the classifier used to compute the coefficients is independent from the one that will be trained using the selected features and both can be changed freely depending on the scenario of interest.

\subsubsection{Selecting k features}
\label{sec:selecting_k_features}
In order to control the selection of a desired number of features, it is possible to include in the objective function a penalty term. This penalty term will be minimized when $k$ variables have value 1. The resulting optimization problem becomes:
\begin{equation*}
	\min \quad y = x^TQx + \left( \sum_{i=1}^{N} x_i - k\right)^2
    \label{eq:kcomb}
\end{equation*}
Note that this penalty term is essential for methods like \ac{MIQUBO} that have a trivial solution where all features are selected, while for the other methods it allows to improve the control on the final outcome.

\subsection{Solving \ac{QUBO} with Traditional Approaches}
The \ac{QUBO} formulation is rather flexible and such problems can be solved with several techniques. Among them, in this work we consider the following: 
\begin{description}
    \item[\acf{SA}:] it is a metaheuristic used to search for solutions of discrete optimization problems \cite{DBLP:journals/science/KirkpatrickGV83}. Based on a temperature parameter decreasing to zero, it searches locally for better solutions while accepting energy-increasing moves with a certain probability that depends on the temperature.
    \item[\acf{TS}:] it is a metahueristic for discrete optimization problems \cite{DBLP:journals/anor/Palubeckis04}. It performs local search and accepts energy increasing moves only if no improving move is available. Moreover, it forbids moves to already visited states. 
    \item[\acf{SD}:] starting from an initial solution, at each step this algorithm performs the variable flip that reduces the energy the most.
\end{description}

\subsection{Solving \ac{QUBO} with Quantum Annealing}
\label{sec:quantum_annealing}

Besides traditional approaches, \ac{QUBO} problems are particularly well suited to be solved via \acf{QA}.
The term \ac{QA}~\cite{Apolloni:192546} originally referred to a meta-heuristic that was proposed as an improvement of \acf{SA} \cite{albash2018adiabatic} for the optimization of objective functions where the quality of a solution is seen as analogous to an \emph{energy}. In particular, \ac{SA} operates by simulating thermal fluctuations and is prone to remain trapped into local optima that are surrounded by a high energy barrier, \idest the worse are the solutions adjacent to the current one, the more difficult for SA is to explore the solution space. 
\ac{QA} was instead designed to leverage quantum tunneling in such a way that even when surrounded by a high energy barrier there is a certain probability to \emph{tunnel} through and escape the local optima. Quantum tunneling is a well-understood quantum mechanical phenomena and, as such, the original proposal for \ac{QA} simulated the process on the available classical systems.

This work instead adopts a different approach to QA, that is to leverage a special-purpose physical device that already exhibits the needed quantum behavior, \idest a Quantum Annealer or \acf{QPU}. In a \ac{QPU} the function to optimize is represented as the energy landscape of a physical system, referred to as the \emph{Hamiltonian} \cite{PhysRevX.6.031015}, and the device acts as a sampler of low-energy solutions.

\subsubsection{Steps to Use a \ac{QPU}}
\label{sec:steps-to-use-QA}
Using a special-purpose device to solve \ac{QUBO} problems brings advantages and disadvantages. The most significant advantage is the very low time-to-solution, as hundreds of solutions can be sampled in few milliseconds. 
As a downside, a physical device has stringent constraints on the size of the problems it can be used to solve. In particular, in a \ac{QPU} each variable is represented with a qubit. The qubits are connected according to a certain topology and only a limited number of connections between qubits exist.
Connections corresponds to entries in the Q matrix of the QUBO problem.
In this paper the D-Wave Advantage \ac{QPU} is used, which has 5600 qubit and a topology called Pegasus, in which every qubit is connected to 15 others \cite{boothby2020next}. This quantum computer easily accessible on the cloud.\footnote{\url{https://www.dwavesys.com/solutions-and-products/cloud-platform/}}

\paragraph{Problem Formulation}
A \ac{QPU} operates by finding the minimums of a certain energy landscape, called Hamiltonian. 
The Hamiltonian is described as a function of the \ac{QUBO} formulation. The \ac{QPU} can use the \ac{QUBO} formulation directly and therefore there is no need for more complex representations as is instead the case for universal quantum computers.

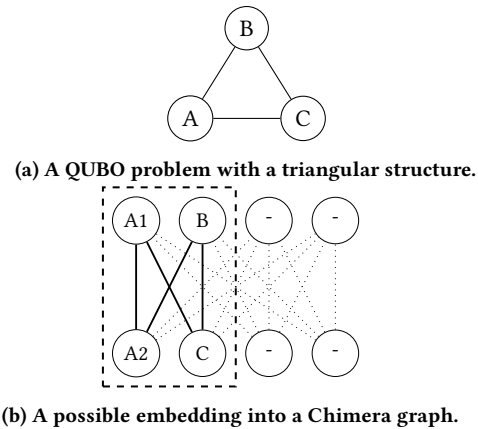
\begin{figure}
     \centering
     \begin{subfigure}[b]{0.5\textwidth}
        \centering
        \begin{tikzpicture}
            \node[shape=circle,draw=black,text=black] (0) at (0,0) {A };
            \node[shape=circle,draw=black,text=black] (1) at (0.75,1.2) {B };
            \node[shape=circle,draw=black,text=black] (2) at (1.5,0) {C };
            
            \path [-] (0) edge (1);
            \path [-] (0) edge (2);
            \path [-] (1) edge (2);
    
        \end{tikzpicture}
        \caption{A \ac{QUBO} problem with a triangular structure.}
        \label{fig:embedding:problem_graph}
        \Description{}
    \end{subfigure}
    \quad
     \begin{subfigure}[b]{0.4\textwidth}
        \centering
        \resizebox{0.50\linewidth}{!}{
        \begin{tikzpicture}[rotate=90,transform shape]
        
            \draw[draw=black,line width=0.3mm, dashed] (-0.5,1.5) rectangle ++(3,2);
        
            \node[shape=circle,draw=black,text=black,minimum size=0.7cm,rotate=-90] (0) at (0,3) {A2};
            \node[shape=circle,draw=black,text=black,minimum size=0.7cm,rotate=-90] (1) at (0,2) {C};
            \node[shape=circle,draw=black,text=black,minimum size=0.7cm,rotate=-90] (2) at (0,1) {-};
            \node[shape=circle,draw=black,text=black,minimum size=0.7cm,rotate=-90] (3) at (0,0) {-};
            \node[shape=circle,draw=black,text=black,minimum size=0.7cm,rotate=-90] (4) at (2,3) {A1};
            \node[shape=circle,draw=black,text=black,minimum size=0.7cm,rotate=-90] (5) at (2,2) {B};
            \node[shape=circle,draw=black,text=black,minimum size=0.7cm,rotate=-90] (6) at (2,1) {-};
            \node[shape=circle,draw=black,text=black,minimum size=0.7cm,rotate=-90] (7) at (2,0) {-};
            \path [thick] (0) edge (4);
            \path [thick] (0) edge (5);
            \path [dotted] (0) edge (6);
            \path [dotted] (0) edge (7);
            \path [thick] (1) edge (4);
            \path [thick] (1) edge (5);
            \path [dotted] (1) edge (6);
            \path [dotted] (1) edge (7);
            \path [dotted] (2) edge (4);
            \path [dotted] (2) edge (5);
            \path [dotted] (2) edge (6);
            \path [dotted] (2) edge (7);
            \path [dotted] (3) edge (4);
            \path [dotted] (3) edge (5);
            \path [dotted] (3) edge (6);
            \path [dotted] (3) edge (7);
        \end{tikzpicture}
        }
        \caption{A possible embedding into a Chimera graph. }
        \label{fig:embedding:unit_cell}
     \end{subfigure}
    \caption{Example of how a problem is embedded in a Quantum Annealer topology. Figure \ref{fig:embedding:problem_graph} shows the structure of a QUBO problem and Figure \ref{fig:embedding:unit_cell} shows a portion of the Chimera topology, an earlier version of Pegasus. Each node represents a qubit and each edge a connection. Each qubit is connected to 4 others of the same cell and to others in different cells. In this case, the triangular problem structure cannot fit directly into the Chimera cell topology therefore the problem variable \emph{A} becomes a logical variable represented by two different qubits \emph{A1} and \emph{A2}. }
    \label{fig:embedding}
    \Description{The figure shows that a triangular problem graph with three variables A, B and C is transformed into a square problem graph. Variable A is duplicated in two variables A1 and A2, connected by an equality constraint, one of them is connected to B and the other one to C.}
\end{figure}

\paragraph{Embedding a \ac{QUBO} problem on the QPU}
When using a \ac{QPU} to solve a \ac{QUBO} problem, the first step is to ensure that the problem can fit on the hardware given its limited connectivity and number of qubits. This step is called \emph{minor embedding} \cite{DBLP:journals/qip/Choi08} and operates by transforming the original \ac{QUBO} problem in an equivalent one that accounts for the \ac{QPU} limited connectivity. This is done, for example, by creating additional variables that will inherit some of the connections of the original ones. Figure \ref{fig:embedding} shows an example of a \ac{QUBO} problem with a triangular structure, that is three variables connected to each other. If such a structure cannot be mapped in the device, the problem structure is transformed into a square by creating an auxiliary variable, therefore using two qubits to represent a single variable. Due to this process, it is not easy to define a relation between the number of \ac{QUBO} problem variables and the number of qubits required by the \ac{QPU}, as the topology of the \ac{QUBO} problem (i.e., the structure of the Q matrix) plays a significant role. This embedding phase can be performed with algorithms that run in polynomial time \cite{DBLP:journals/qip/Choi08}.

\paragraph{Sampling Solutions}
Once the problem has been embedded, it is possible to submit the problem to the \ac{QPU} via APIs. The device operates by \emph{sampling} low-energy solutions by repeating the physical \ac{QA} process multiple times for the desired length of time. Depending on the problem, it may be more ore less likely that a good solution is sampled, therefore it is common to sample a significant number of solutions, typical values are $10^2 - 10^4$. Each solution is associated to the corresponding energy, which can be used to select the desired ones.

\paragraph{Advanced Controls}
The \ac{QPU} has several settings that can be used to control the underlying physical process, such as the duration of the annealing process that returns a single sample (typical durations are between $1-100 \mu s$), the schedule of the annealing process, time offsets for different qubits etc. Finding the optimal setup of the physical process is a complex task that is yet not well-understood and therefore constitutes an open research question that goes beyond the scope of this work.

In this work, we explore two state-of-the-art approaches to \ac{QA}:
\begin{description}
    \item[Advantage Quantum Annealer (QPU):] uses a real \ac{QPU} to search for solutions, with default hyperparameters (in particular, annealing time of $20\mu s$).
    \item[D-wave Leap Hybrid (Hybrid):] uses a hybrid quantum-classical approach to decompose the \ac{QUBO} problem and partially solve smaller problems directly on the annealer; available as a cloud service from D-Wave Leap\footnote{D-Wave Leap - \url{https://cloud.dwavesys.com/leap/}}.
\end{description}

\section{Experimental Pipeline}
\label{sec:exprimental-pipeline}
The effectiveness of the \ac{QPU} is assessed on two different tasks, classification and ranking. This section presents the tasks, the used datasets and algorithms as well as the details of the experimental pipeline. The source code to reproduce these experiments is publicly available online. \footnote{\url{https://github.com/qcpolimi/SIGIR22_QuantumFeatureSelection.git}}

\subsection{Classification Task}
The classification task consists in classifying a sample from a dataset (such as an image, a sound, a sentence) into two or more different classes, based on the sample's features.
These features can be heterogeneous (numerical, categorical) and may refer to a direct measurement (\eg the size of an object) or may be extracted from other data (\eg the number of times a certain word appears in a text).
Given the usually high number of features, feature selection is important and largely used in classification, in order to identify the most useful subset of features for the specific task.

\paragraph{Dataset} 
For the classification experiments, datasets are taken from OpenML~\cite{OpenML2013}, with a number of features ranging from 34 to 5000. The datasets come from heterogeneous fields and tasks, among which: clinical diagnosis (\texttt{\texttt{waveform-5000}}, \texttt{SPECTF}), organic chemistry (\texttt{Bioresponse}), forest cover type (\texttt{covertype}), duplicate locations resolution (\texttt{nomao}), digit recognition (\texttt{USPS}, \texttt{SVHN\_small}, \texttt{gisette}), e-mail filtering (\texttt{spambase}). Some datasets have features extracted from images (\texttt{SPECTF}, \texttt{SVHN\_small}, \texttt{USPS}, \texttt{gisette}), signals (\texttt{waveform-5000}), e-mail text (\texttt{spambase}) and even sound (\texttt{isolet}). The number of samples spans from a couple hundreds to tens of thousand, with one large dataset of hundreds of thousands samples (\texttt{covertype}). Almost every dataset has two classes, except from \texttt{waveform-5000} (3), \texttt{USPS} and \texttt{SVHN\_small} (10) and \texttt{isolet} (26).

\paragraph{Classification algorithm} The algorithm used for the classification task is Random Forest, which is a widely used ensemble method that builds multiple decision trees. The class is determined by voting as the one selected by the most random trees. Random Forest has been chosen because it is able to provide good classification accuracy with limited fine-tuning and generally works well across multiple tasks and domains, which makes it well-suited for this study. The optimized metric is classification accuracy.

\paragraph{Data split} 
The dataset is randomly split in two sets, training and testing, respectively with 70\% and 30\% of the samples. The splits are generated by randomly sampling data samples but ensuring the approximate class distribution is maintained. This is especially important for datasets that exhibit class imbalance. Due to the limited size of some datasets, 5-fold Cross Validation is applied on the training data instead of creating a further split for validation.

\subsection{Ranking Task}
To evaluate the different feature selection strategies in a traditional \ac{IR} setting, we consider the \ac{LtR} task. The \ac{LtR} task requires sorting a set of documents according to their expected relevance for a given query. For each query document pair, a set of features is computed.
Additionally, for the pairs query document used during the training phase, we have the relevance judgments describing whether the document is relevant to the query. An \ac{LtR} approach often consists in training a model, using relevance judgments as supervision, that uses features of query document pairs to predict the document relevance to the query. This, in turn, allows sorting the documents according to the predicted relevance for a query.   

\paragraph{Dataset} Following previous literature on \ac{LtR}, we adopt LETOR to assess the performance of different feature selectors. 
LETOR is a collection of datasets explicitly meant for the \ac{LtR} task. Currently, there are two main versions, LETOR 3.0~\cite{QinEtAl2010} and LETOR 4.0~\cite{QinLiu2013} that incorporate respectively seven and two distinct datasets.

LETOR 3.0 includes seven different datasets, among which in our experiments we consider \texttt{OHSUMED}. The corpus of this dataset contains more than 300,000 scientific publications (title and abstract of medical papers). \texttt{OHSUMED} includes 45 features. 
Relevance in the \texttt{OHSUMED} collection is ternary and can be either 0, 1, or 2. Such values are assigned respectively to non-relevant, partially relevant, and highly relevant documents.

LETOR 4.0 contains the \texttt{MQ2007} and \texttt{MQ2008} datasets. They are built upon TREC Million Queries track from the years 2007 and 2008 and have 1700 and 800 queries each. \texttt{MQ2007} and \texttt{MQ2008} are based on the \texttt{.gov2} collection that consists of more than 25 million documents.
Akin to \texttt{OHSUMED}, relevance judgements for both \texttt{MQ2007} and \texttt{MQ2008} are ternary.
For each pair query document to be reranked, LETOR 4.0 contains 46 features. 

\paragraph{Ranking algorithm} LambdaMART~\cite{Burges2010} is a state-of-the-art \ac{LtR} approach that combines the LambdaRank~\cite{BurgesEtAl2006} optimization algorithm and Multiple Additive Regression Trees (MART)~\cite{Friedman2001}.
We adopt the implementation of LambdaMART available in RankLib\footnote{https://sourceforge.net/p/lemur/wiki/RankLib/}, with default hyperparameters. We used nDCG with a cutoff at ten as optimization measure, as commonly done in the literature.

\paragraph{Data split} Given the cost of selecting the features using quantum-based strategies, we limit our analysis to a single collection split. In particular, for each dataset, we use 60\% of the queries as a training set --  both select the features and train the classifiers. The remaining 40\% of the queries in each dataset are split evenly in validation and test sets: the former allows selecting the optimal number of features, and the latter serves to evaluate the performance of the classifiers.

\subsection{Selecting the optimal set of features}
Feature selection is performed by applying to the training data \ac{QUBO} and the baseline linear feature selection methods described in Section \ref{sec:related}.
Note that all methods require to specify the number of features to be selected, $k$, which constitutes a hyperparameter.
How the $k$ features are selected differs between linear filter methods and \ac{QUBO}-based methods. For linear filter methods the $k$ features selected are the ones with the highest computed scores. For \ac{QUBO} methods a component is added to the loss function as described in Section \ref{sec:selecting_k_features}.
The search range for $k$ depends on the number of features $|F|$ of the dataset.
If there are less than 50 features, $k$ is selected from a range between 1 and $|F|$ with a unitary step.
Otherwise, a maximum of 50 values for $k$ are explored, equally distributed between 1 and $|F|$ excluded.

For each of the explored $k$ values, the resulting selected features are used to train either the classifier or the ranker, depending on the task, and the resulting model is evaluated on the validation data. 
Note that the number $N$ of selected features may be different from $k$, the target number of features to be selected, because the target value $k$ is expressed as a soft constraint.
The $N$ value associated to the features with the best model quality on the validation data is selected and a final model is trained on the union of training and validation data. The quality of this model evaluated on the test data is then reported.

For all QUBO solvers 100 solutions are generated or sampled, except for the Leap Hybrid method that only returns the best solution found according to its energy.

\begin{table*}[h!]
    \caption{Classification accuracy for all \ac{QUBO} feature selection methods. The first column refers to all dataset features, the other columns refer to the various solvers used to solve the \ac{QUBO} feature selection problem. Superscripts $^A$, $^D$ and $^T$ indicate statistical difference between the Quantum Solvers and Simulated Annealing, Steepest Descent and Tabu Search respectively, determined using the procedure described in Section \ref{sec:results}. The best results for each \ac{QUBO} method and dataset are highlighted in bold. Results are missing for the \ac{QPU} when the problem required more qubits than the available ones, in such instances the only Quantum Solver that could be used is Hybrid. 
    }
    \label{tab:QUBO-classification-all-solvers}
    \footnotesize
    \centering
    \begin{tabular}{cl|rc|rl|rl|rc|rc|rc}
    \toprule
    \multirow{3}{*}{Method} & 
    \multirow{3}{*}{Dataset} &
    \multicolumn{2}{c}{All Features} \vline & \multicolumn{4}{c}{Quantum Solver} \vline & \multicolumn{6}{c}{Traditional Solver}\\
    & & & & \multicolumn{2}{c}{QPU}  & \multicolumn{2}{c}{Hybrid}  \vline & \multicolumn{2}{c}{SA}  & \multicolumn{2}{c}{SD} & \multicolumn{2}{c}{TS} \\
    & &            F & Accuracy &   N & Accuracy &         N & Accuracy &                  N & Accuracy &               N & Accuracy &           N & Accuracy \\
    \midrule
    \multirow{11}{*}{\shortstack{\ac{QUBO}\\Correlation}}
    & waveform-5000 &           40 &   0.8540 &  31             &   0.8393 &        36 &   0.8333 &                 36 &   0.8320 &              34 &   0.8333 &          40 &   \textbf{0.8593} \\
    & SPECTF        &           44 &   \textbf{0.8667} &  14    &   0.8476 &        16 &   0.8286 &                 18 &   0.8286 &              16 &   0.8381 &          15 &   0.8476 \\
    & covertype     &           54 &   0.9605 &  46             &  0.9526$^{ADT}$ &        43 &   0.9352$^T$ &                 41 &   0.9347 &              43 &   0.9342 &          49 &   \textbf{0.9646} \\
    & spambase      &           57 &   0.9631 &  47             &  0.9573 &        47 &   0.9594 &                 51 &   0.9602 &              50 &   0.9587 &          56 &   \textbf{0.9660} \\
    & nomao         &          118 &   0.9654 &  93             &   \textbf{0.9659} &       104 &   0.9649 &                110 &   0.9654 &             111 &   0.9650 &         118 &   0.9654 \\
    & tecator       &          124 &   0.9167 &  67             &   \textbf{0.9306} &        26 &   \textbf{0.9306} &                 26 &   0.9028 &              35 &   0.9167 &          79 &   0.9167 \\
    & USPS          &          256 &   \textbf{0.9642} &        &    \phantom{0.0}- &       198 &   0.9599 &                201 &   0.9613 &             204 &   0.9624 &         253 &   0.9599 \\
    & isolet        &          617 &   \textbf{0.9432} &        &    \phantom{0.0}- &       565 &   0.9402 &                565 &   0.9406 &             556 &   0.9402 &         617 &   0.9406 \\
    & Bioresponse   &         1776 &   0.7913 &                 &    \phantom{0.0}-  &        76 &   0.7940 &                 76 &   0.7948 &             403 &   \textbf{0.8002} &         387 &   0.7984 \\
    & SVHN\_small    &         3072 &   \textbf{0.5801} &       &    \phantom{0.0}-  &       401 &   0.5170$^T$ &                401 &   0.5203 &             352 &   0.5166 &        1468 &   0.5619 \\
    & gisette       &         5000 &   0.9676 &                 &    \phantom{0.0}-  &      4536 &   0.9471$^T$ &               4536 &   0.9486 &            4536 &   0.9462 &        2560 &   \textbf{0.9681} \\
    \midrule
    \multirow{11}{*}{\shortstack{\ac{QUBO}\\Boosting}}
    & waveform-5000 &           40 &   0.8540 &  23 &   0.8287 &        24 &   \textbf{0.8620} &                 23 &   0.8567 &              25 &   0.8520 &          28 &   0.8567 \\
    & SPECTF        &           44 &   0.8667 &   6 &   0.8286 &        30 &   \textbf{0.8857} &                 29 &   0.8571 &              26 &   0.8286 &          13 &   0.8571 \\
    & covertype     &           54 &   0.9605 &  33 &   0.9621$^{ADT}$ &        31 &   0.9667$^{ADT}$ &                 30 &   \textbf{0.9697} &              33 &   0.9643 &          25 &   0.9650 \\
    & spambase      &           57 &   \textbf{0.9631} &  37 &   0.9558 &        35 &   0.9529 &                 37 &   0.9573 &              36 &   0.9566 &          53 &   0.9587 \\
    & nomao         &          118 &   0.9654 &  64 &   0.9655 &        79 &   0.9645 &                 79 &   0.9641 &              76 &   0.9650 &         112 &   \textbf{0.9667} \\
    & tecator       &          124 &   0.9167 &  69 &   0.9167 &        23 &   0.9167 &                 14 &   0.9167 &              40 &   0.8889 &          16 &   \textbf{0.9444} \\
    & USPS          &          256 &   \textbf{0.9642} &     &        \phantom{0.0}-  &       161 &   0.9573 &                161 &   0.9613 &             165 &   0.9591 &         234 &   0.9624 \\
    & isolet        &          617 &   \textbf{0.9432} &     &        \phantom{0.0}-  &       467 &   0.9380 &                495 &   0.9415 &             495 &   0.9406 &         604 &   0.9380 \\
    & Bioresponse   &         1776 &   0.7913 &     &         \phantom{0.0}- &      1229 &   0.7922 &                997 &   0.7940 &            1345 &   0.7948 &        1123 &   \textbf{0.7993} \\
    & SVHN\_small    &         3072 &   0.5801 &     &        \phantom{0.0}-  &      2245 &   0.5670 &               1855 &   0.5623 &            2099 &   \textbf{0.5831} &        1469 &   0.5619 \\
    & gisette       &         5000 &   0.9676 &     &        \phantom{0.0}-  &      1750 &   0.9729 &               2079 &   \textbf{0.9752} &            1002 &   0.9733 &        2558 &   0.9690 \\
    \midrule
    \multirow{11}{*}{\ac{MIQUBO}}
    & waveform-5000 &           40 &   0.8540 &  20 &   0.8547 &        20 &   0.8553 &                 27 &   \textbf{0.8580} &              38 &   0.8540 &          40 &   0.8493 \\
    & SPECTF        &           44 &   0.8667 &  44 &   \textbf{0.8857} &        44 &   \textbf{0.8857} &                 44 &   \textbf{0.8857} &              44 &   0.8476 &          44 &   0.8476 \\
    & covertype     &           54 &   0.9605 &  50 &  \textbf{0.9646}$^{AT}$ &        49 &   0.9616$^{DT}$ &                 36 &   0.9624 &              27 &   0.9642 &          54 &   0.9605 \\
    & spambase      &           57 &   0.9631 &  52 &   0.9638 &        53 &   0.9616 &                 52 &   0.9645 &              54 &   0.9638 &          57 &   \textbf{0.9660} \\
    & nomao         &          118 &   0.9654 &  96 &   \textbf{0.9667} &       109 &   0.9656 &                118 &   0.9656 &             113 &   0.9661 &         118 &   0.9656 \\
    & tecator       &          124 &   \textbf{0.9167} &  67 &   \textbf{0.9167} &         8 &   \textbf{0.9167} &                 42 &   \textbf{0.9167} &              19 &   \textbf{0.9167} &         124 &   \textbf{0.9167} \\
    & USPS          &          256 &   0.9642 &     &        \phantom{0.0}-  &       219 &   0.9620 &                256 &   0.9624 &             185 &   \textbf{0.9649} &         256 &   0.9634 \\
    & isolet        &          617 &   0.9432 &     &        \phantom{0.0}-  &       501 &   \textbf{0.9436} &                617 &   0.9389 &             600 &   0.9410 &         617 &   0.9427 \\
    & Bioresponse   &         1776 &   0.7913 &     &        \phantom{0.0}-  &      1709 &   \textbf{0.7966} &               1419 &   0.7922 &            1492 &   0.7869 &        1776 &   0.7922 \\
    & SVHN\_small    &         3072 &   0.5801 &     &       \phantom{0.0}-   &      1871 &   0.5787 &               1337 &   0.5787 &            2136 &   \textbf{0.5911} &        1609 &   0.5697 \\
    & gisette       &         5000 &   0.9676 &     &        \phantom{0.0}-  &      1076 &   0.9743 &               1382 &   \textbf{0.9771} &            2705 &   0.9710 &        2598 &   0.9686 \\

\bottomrule
\end{tabular}

\end{table*}

\begin{table*}[h]
    \caption{NDCG at 10 on the Ranking task for all \ac{QUBO} feature selection methods. The first column refers to all dataset features, the other columns refer to the various solvers used to solve the \ac{QUBO} feature selection problem. The best results for each \ac{QUBO} method and dataset are highlighted in bold. 
    }
    \label{tab:QUBO-ranking-all-solvers}
    \footnotesize
    \centering\begin{tabular}{cl|rr|rr|rr|rr|rr|rr}
    \toprule
    \multirow{3}{*}{Method} & 
    \multirow{3}{*}{Dataset} &
    \multicolumn{2}{c}{All Features} \vline & \multicolumn{4}{c}{Quantum Solver} \vline & \multicolumn{6}{c}{Traditional Solver}\\
    & & & & \multicolumn{2}{c}{QPU}  & \multicolumn{2}{c}{Hybrid}  \vline & \multicolumn{2}{c}{SA}  & \multicolumn{2}{c}{SD} & \multicolumn{2}{c}{TS} \\
    & &            F &             NDCG &   N &   NDCG &      N &   NDCG &   N &             NDCG &   N &   NDCG &    N &             NDCG \\
    \midrule
    \multirow{3}{*}{\shortstack{\ac{QUBO}\\Correlation}}
    & OHSUMED &           45 &           0.3882 &  35 & 0.3706 &     44 & 0.3659 &  44 &  \textbf{0.3903} &  44 & 0.3872 &   26 &           0.3332 \\
    & MQ2007  &           46 &           0.4721 &  43 & 0.4731 &     39 & 0.4720 &  39 &           0.4738 &  28 & 0.4733 &   24 &  \textbf{0.4757} \\
    & MQ2008  &           46 &  \textbf{0.4891} &   4 & 0.4831 &     21 & 0.4677 &  21 &           0.4657 &  19 & 0.4854 &   25 &           0.4609 \\
    \midrule
    \multirow{3}{*}{\shortstack{\ac{QUBO}\\Boosting}}
    & OHSUMED &           45 &           0.3882 &   9 &           0.3457 &     32 & 0.3632 &  19 & 0.3382 &  40 &  \textbf{0.4002} &   23 & 0.3884 \\
    & MQ2007  &           46 &           0.4721 &  43 &  \textbf{0.4760} &     37 & 0.4638 &  42 & 0.4632 &  36 &           0.4640 &   35 & 0.4662 \\
    & MQ2008  &           46 &  \textbf{0.4891} &   8 &           0.4599 &      8 & 0.4736 &  20 & 0.4852 &  39 &           0.4759 &   34 & 0.4853 \\
    \midrule
    \multirow{3}{*}{MIQUBO}
    & OHSUMED &           45 &           0.3882 &  11 &           0.3685 &     17 & 0.3750 &  17 &  \textbf{0.3942} &   6 & 0.3755 &    4 &           0.3882 \\
    & MQ2007  &           46 &           0.4721 &  25 &  \textbf{0.4798} &     34 & 0.4722 &  34 &           0.4685 &  34 & 0.4722 &    2 &           0.4721 \\
    & MQ2008  &           46 &  \textbf{0.4891} &   1 &           0.4743 &     18 & 0.4791 &  18 &           0.4791 &  18 & 0.4791 &   18 &  \textbf{0.4891} \\
    \bottomrule
    \end{tabular}
\end{table*}

\section{Results}
\label{sec:results}
This section presents the results aiming to assess the capability of the \ac{QPU} to solve \ac{QUBO} problems competitively, by comparing the solutions obtained by all \ac{QUBO} solvers both in terms of effectiveness and efficiency. Overall, more than 10k experiments were conducted, of which 900 on the \ac{QPU} directly and 1650 using the Hybrid solver.

\begin{table}[h]
    \caption{Effectiveness of the baseline linear feature selection methods on the Classification and Ranking tasks, measured respectively with classification accuracy and NDCG at 10. Results that are better than or equal to the ones using all the features or QUBO methods are highlighted in bold. 
    }
    \label{tab:QUBOall-baselines}
    \footnotesize
    \centering
    \begin{tabular}{l|cccccc}
\toprule
\begin{tabular}{@{}c@{}}Dataset\end{tabular} &  
\begin{tabular}{@{}c@{}}ANOVA\end{tabular} &
\begin{tabular}{@{}c@{}}Chi2\\Test\end{tabular} &
\begin{tabular}{@{}c@{}}MI\end{tabular} & 
\begin{tabular}{@{}c@{}}Pearson\\Corr.\end{tabular} &  
\begin{tabular}{@{}c@{}}Linear\\Boost.\end{tabular} &  
\begin{tabular}{@{}c@{}}Variance\\Thr.\end{tabular}  \\
\midrule
waveform-5000             & 0.8593 & 0.6893 & 0.8573 &   0.8473 &     0.8533 &  0.8567 \\
SPECTF                    & 0.8857 & 0.8571 & 0.8190 &   0.8762 &     0.8762 &  0.8571 \\
covertype                 & 0.9644 & 0.9642 & 0.9612 &   0.9610 &     \textbf{0.9678} &  0.9655 \\
spambase                  & 0.9616 & 0.9638 & 0.9631 &   0.9529 &     \textbf{0.9681} &  0.9616 \\
nomao                     & 0.9654 & 0.9639 & 0.9656 &   0.9646 &     0.9652 &  0.9648 \\
tecator                   & 0.9306 & 0.8472 & 0.9028 &   0.9028 &     0.9306 &  0.9167 \\
USPS                      & 0.9616 & 0.9616 & 0.9624 &   0.9627 &     0.9631 &  \textbf{0.9652} \\
isolet                    & 0.9372 & 0.9406 & 0.9406 &   0.9432 &     0.9389 &  \textbf{0.9444} \\
Bioresponse               & 0.7966 & 0.7948 & 0.7824 &   0.7851 &     0.7922 &  0.7975 \\
SVHN\_small                & 0.5747 & 0.5770 & 0.5606 &   0.5596 &     0.5636 &  0.5616 \\
gisette                   & 0.9748 & 0.9181 & 0.9748 &   0.9700 &     0.9748 &  0.9710 \\
\midrule
OHSUMED &           0.3639 & 0.3475 & 0.3582 & 0.3639 & 0.3900 & 0.3470 \\
MQ2007  &           0.4714 & 0.4614 & 0.4722 & 0.4696 & 0.4703 & 0.4693 \\
MQ2008  &           0.4853 & 0.4788 & 0.4806 & 0.4853 & 0.4831 & 0.4748 \\
\bottomrule
\end{tabular}

\end{table}
    
\subsection{Effectiveness} 
The results obtained by the QUBO solvers are compared and displayed in Table \ref{tab:QUBO-classification-all-solvers} (Classification Task) and Table \ref{tab:QUBO-ranking-all-solvers} (Ranking Task).

To determine statistically significant differences between features selection strategies applied to the classification task, we employ the McNemar's test with significance level $\alpha=0.05$ and Bonferroni correction following the procedure described by~\citet{japkowiczShah2011}. As a general trend, quantum solvers perform as well as traditional ones, as expected. The only exception to this pattern is represented by the \texttt{covertype} dataset, where, in several cases, there are statistically significant differences between different approaches. It should be noted that, even for \texttt{covertype}, there is no dominance of traditional solvers over quantum ones, nor vice versa: the best performing solver is linked to the chosen heuristics.

Concerning the ranking task, we assess the presence of statistically significant differences between selection strategies using a two-way ANOVA, with topic and feature selector factors, followed by Tukey's posthoc pairwise comparison procedure with significance level $\alpha=0.05$. Such a comparison strategy follows the one proposed originally by~\cite{tague1995statistical}, where the system factor is replaced with the feature selector one. 
On all collections, the statistical procedure does not deem any selector to be statistically better than others. This indicates overall comparable performance when using either traditional or quantum approaches to the solution of the QUBO problem: using quantum strategies, we can expect results that are at least as good as those that we would have achieved otherwise.

The first observation that can be made is that there is no single \ac{QUBO} solver that is superior to the others, rather, the solver that is able to achieve the best result is different depending on the dataset. Across all experiments TS is able to reach the best result 11 times, SA and \ac{QPU} 8 times, Hybrid 7 times and SD 6 times.
This is likely due to the peculiarities of each dataset, task and, to some extent, the stochastic nature of some solvers. Some differences instead emerge by comparing across tasks, in particular no feature selection approach is able to improve the effectiveness on dataset \texttt{MQ2008} and the Hybrid solver is slightly less effective when applied to the Ranking task.
The behavior of the various \ac{QUBO} solvers remains instead consistent across the feature selection methods, although different \ac{QUBO} heuristics result in different overall effectiveness. For example, \ac{MIQUBO} appears to produce better results compared to the others.

Looking at the \ac{QPU} solver, its effectiveness is very close, if not almost identical, to that of the other solvers, sometimes resulting in the best selection of features.
In very few cases the solution obtained with the \ac{QPU} is worse compared to the other solvers, but within 5\% of the best one. This result indicates that the \ac{QPU} is indeed a reliable solver that can be used to tackle real problems across different datasets, heuristics and tasks.
Note however that the largest problem that could be solved directly on the \ac{QPU} had 124 features. Although the \ac{QPU} has more than 5000 qubits, the \ac{QUBO} matrix resulting from the feature-selection problem is fully-connected and therefore its structure is difficult to fit on the limited connectivity structure of the \ac{QPU}, therefore the problem size that can fit the hardware is greatly reduced.
For larger problems it is still possible to use the Hybrid QPU-Classical approach, which was used for datasets of up to 5000 features but is able to tackle even larger problems.

In terms of the number of selected features, again no clear pattern emerges with no solver able to consistently provide better solutions with a lower number of selected features. Comparing the number of selected features with the desired number, \idest $k$, reveals that in half of all cases the number of actually selected features is within 10\% of $k$, while the remaining half is split equally in cases where the number of selected features is lower and higher. Since the desired number of features is a penalty, as described in subsection \ref{sec:selecting_k_features}, increasing its strength in the final QUBO problem will allow to better control the resulting number of selected features.

Lastly, Table \ref{tab:QUBOall-baselines} compares the effectiveness of simple linear feature selection baselines. Such baselines are generally unable to provide better solutions when compared with the QUBO models and the baseline using all features. For the classification task SVC Boosting and Variance Threshold are able to provide better solutions in two cases each, but for different datasets, while for the ranking task none of the baselines exhibits better solution effectiveness.

\begin{table*}[h]
    \caption{Total time (in seconds) required to solve the $k$ QUBO models with different QUBO solvers, for both the Classification and the Ranking tasks. Due to space limitations, only two of the three methods are reported. However, QUBO-Boosting behaves similarly to QUBO-Correlation, and the complete results can be found in the online material. Results are missing for the QPU when the problem required more qubits than the available ones, in such instances the only Quantum Solver that could be used is Hybrid.
    }
    \label{tab:QUBO-solution-time}
    \footnotesize
    \centering
    \begin{tabular}{l|r|rr|rrr|rr|rrr}
\toprule
\multirow{3}{*}{Dataset} & 
\multirow{3}{*}{F} &
\multicolumn{5}{c}{QUBO-Correlation} \vline & \multicolumn{5}{c}{MIQUBO} \\
& & \multicolumn{2}{c}{Quantum Solver} \vline& \multicolumn{3}{c}{Traditional Solver}  \vline & \multicolumn{2}{c}{Quantum Solver} \vline & 
\multicolumn{3}{c}{Traditional Solver} \\
& &             QPU & Hybrid &       SA &      SD &    TS &                   QPU & Hybrid &      SA &      SD &    TS \\
\midrule
waveform-5000             &    40 &            39.0 &     394.4 &     57.9 &     7.8 &   474.6 &                  39.2 &     385.9 &     7.5 &     1.0 &    78.6 \\
SPECTF                    &    44 &            48.4 &     426.3 &     72.3 &     9.3 &   522.4 &                  45.1 &     421.7 &     9.5 &     1.2 &    86.7 \\
covertype                 &    54 &           130.1 &     491.5 &     17.0 &     1.9 &    99.1 &                 127.3 &     809.9 &    17.0 &     1.9 &    99.0 \\
spambase                  &    57 &            65.9 &     487.4 &    107.1 &    15.7 &   598.9 &                  51.5 &     486.4 &    17.6 &     2.0 &    99.1 \\
nomao                     &   118 &           100.1 &     507.5 &    323.6 &    54.3 &   620.9 &                  98.1 &     512.5 &    46.6 &     7.9 &   101.9 \\
tecator                   &   124 &            95.6 &     504.3 &    374.2 &    57.7 &   623.5 &                  90.2 &     515.1 &    62.3 &     8.8 &   102.3 \\
USPS                      &   256 &                - &     541.1 &   1174.0 &   220.4 &   785.2 &                      - &     542.6 &   138.0 &    34.6 &   124.3 \\
isolet                    &   617 &                - &     640.2 &   6280.7 &  1260.4 &  1735.7 &                      - &     642.4 &   838.1 &   201.6 &   263.8 \\
Bioresponse               &  1776 &                - &    1435.1 &  53978.2 & 10814.2 & 10531.4 &                      - &    1423.4 & 11031.8 &  1727.8 &  1632.8 \\
SVHN\_small                &  3072 &                - &    3575.4 & 124041.7 & 33802.1 & 32440.0 &                      - &    3591.1 & 23332.5 &  5455.9 &  5146.8 \\
gisette                   &  5000 &                - &    8666.3 &  80062.4 & 18811.6 & 18238.1 &                      - &    8606.4 & 78635.7 & 14210.1 & 14610.5 \\
\midrule
OHSUMED         &  45 &            154.1 &     436.8 & 13.5 & 1.7 &  89.4 &                 497.9 &     435.1 & 13.9 & 1.5 &  89.2 \\
MQ2007          &  46 &            472.5 &     444.5 & 15.6 & 1.9 &  91.5 &                  74.0 &     445.0 & 12.8 & 1.5 &  91.2 \\
MQ2008          &  46 &            122.7 &     436.7 & 15.0 & 1.6 &  91.7 &                  96.4 &     443.1 & 13.0 & 1.5 &  91.3 \\
\bottomrule
\end{tabular}

\end{table*}

\begin{table}[h]
    \caption{Drill down of the time (in seconds) required to solve the QUBO models generated with MIQUBO. The other heuristics behave similarly. The Embedding column refers to the time required to embed the problem on the QPU. The columns under QPU show the time-to-solution as observed by the local client (Total), which corresponds to the QPU time reported in \tabref{tab:QUBO-solution-time}, splitted between the actual physical annealing process (Sampling) and the latency due to the data transfer as well as further waiting time after the task is queued (Latency). Note that the Latency time is more than one order of magnitude higher than the Sampling time.
    }
    \label{tab:QPU-timing-drilldown}
    \footnotesize
    \centering
    
    \begin{tabular}{l|r|r|rcr}
\toprule
\multirow{2}{*}{Dataset} & 
\multirow{2}{*}{F} & 
\multirow{2}{*}{Embedding} & 
\multicolumn{3}{c}{QPU} \\
 &   &   &   Total &  Sampling &  Latency  \\
\midrule
waveform-5000             &   40 &        7.0 &  39.2 &       0.9 &     38.3 \\
SPECTF                    &   44 &        5.3 &  45.1 &       1.1 &     44.0 \\
covertype                 &   54 &        7.9 & 127.3 &       1.2 &    126.1 \\
spambase                  &   57 &       14.2 &  51.5 &       1.0 &     50.5 \\
nomao                     &  118 &      209.9 &  98.1 &       1.6 &     96.5 \\
tecator                   &  124 &      164.6 &  90.2 &       1.5 &     88.7 \\
\midrule
OHSUMED                     &  45 &        7.2 & 497.9 &       1.0 &    496.9 \\
MQ2007                      &  46 &       17.0 &  74.0 &       1.0 &     72.9 \\
MQ2008                      &  46 &       10.3 &  96.4 &       1.2 &     95.3 \\
\bottomrule
\end{tabular}

\end{table}

\subsection{Efficiency} 
This section discusses the computational cost of each step required to use the \ac{QPU}, following the same structure as Section \ref{sec:steps-to-use-QA}.
The experiments for all Quantum and Classical solvers have been conducted on the same machine to ensure the computational time is comparable and measure the time-to-solution as observed by the local client. For the Traditional Solvers, the time-to-solution of the QUBO problem corresponds to the actual time spent by the solver in finding a solution. For the Quantum Solvers, instead, the time-to-solution has three components: Embedding $+$ Latency $+$ Sampling, as detailed in Table \ref{tab:QPU-timing-drilldown}. Due to the characteristics of the embedding process, as will be described in this section, for the \ac{QPU} Solver we consider the time-to-solution as only Latency $+$ Sampling, keeping the Embedding time separate.

\paragraph{Problem Formulation}
In order to allow a fair comparison of the time-to-solution of the \ac{QUBO} solvers one first has to account for the time required to compute the heuristics used by the \ac{QUBO} problem itself. 
This is a computationally expensive step that depends both on the number of features and the number of samples in the dataset, roughly requiring to compute $|F|^2/2$ coefficients. This process requires a few seconds until the number of features exceeds 600. For example, computing the \ac{QUBO} coefficients requires a maximum of ten seconds for \texttt{USPS}, while it requires between 4 minutes and one hour for \texttt{gisette}, depending on the heuristic. The time required to compute the \ac{QUBO} coefficients must be taken into account to ensure the time required to solve the problem by the various solvers is put in the right perspective. It should be noted however that the \ac{QUBO} coefficients must be computed only once and then it can be reused, \eg deriving the \ac{QUBO} for each $k$ value, to search for different hyperparameters for the \ac{QUBO} problem or to for the solver etc. Overall, even though in some cases the time required to generate the \ac{QUBO} coefficients is quite long, it is always lower than the time required to solve it. 

\paragraph{Embedding on the QPU}
Table \ref{tab:QPU-timing-drilldown} shows the detail of the computational time for the \ac{QPU}. First, the time required for the minor embedding process varies greatly and while it is in the range of seconds for smaller problems, it can be up to two minutes for problems of a size close to the maximum that can fit on the QPU. Although fully-connected problems are the most difficult and slowest to embed, the embedding only depends on the problem \emph{structure}, not on the specific coefficients. This means that the embedding for a fully-connected \ac{QUBO} problem of a certain number of variables can be computed once and then it can be reused for any other fully-connected \ac{QUBO} problem.

\paragraph{Sampling Solutions}
Table \ref{tab:QUBO-solution-time}, instead, compares the computational time required by all solvers for the \ac{MIQUBO} and QUBO-Correlation problems. QUBO-Boosting is omitted for space reasons as it behaves very similarly to QUBO-Correlation. The full results are available in the online material. 
For both QUBO-Correlation and QUBO-Boosting the \ac{QPU} has a solution time always lower than both SA and TS, with only SD being faster. In particular, for problems close to the maximum size that can fit on the \ac{QPU}, SA has a solution time of three times that of the \ac{QPU} while TS six times. For larger problems, the time gap between SD and TS reduces drastically and Hybrid becomes the fastest solver. It is interesting to notice how for the largest problem of 5000 features SA requires ten times longer than Hybrid.
MIQUBO instead shows a different behaviour, with SA comparably much faster and able to show better performance than the \ac{QPU} for smaller problems. As the problem size grows however the solvers show a similar behaviour as that of QUBO-Correlation, with Hybrid becoming the fastest solver, SA and TS close at twice its solution time and SA being the slowest.

As a general observation it can be seen how the computational time grows as the number of features increases, as can be expected due to the increased problem complexity and values of $k$ explored. The main reason for the increase is however different for classical solvers and the QPU. For classical solvers, the increased solution time is due essentially the greater problem complexity. The \ac{QPU} instead has a fixed annealing time such that the solution is always returned in constant time for any problem that can fit on its hardware. In these experiments the increased computational time is essentially due to the data transfer, since the coefficients are transmitted to the \ac{QPU} over the global network and bigger problems will require to transfer more data. Similarly, the Hybrid solver too has a time limit which is a function of the problem size. Note that the constant solution time by the \ac{QPU} does not offer guarantees regarding its solution quality. For more complex problems it may be needed to sample an increasing number of solutions or to increase the duration of the annealing process. In general, there is no agreed rule on how to choose those settings that impact the underlying physical process. Some bounds have been proven rigorously but only for ideal devices and are, in practice, not applicable to the available \acp{QPU} \cite{McGeoch}.

Another important aspect apparent from Table \ref{tab:QPU-timing-drilldown} is the large difference between the actual duration of the physical annealing process and the time spent by the client waiting for the response. This constitutes another limitation of the current way a \ac{QPU} can be used as it creates a very large overhead due to the high-latency transfer of the problem data and solution through the global internet network and includes possible waiting time until the \ac{QPU} becomes available.

Overall, most of the time currently required to solve a fully-connected \ac{QUBO} problem with a \ac{QPU} can be eliminated by offering pre-built embeddings and low-latency access.

\section{Discussion and Future Works}
\label{sec:discussion}
This work shows the application of currently available Quantum Annealer technology to the feature selection task in classification and ranking problems. 
The results show that: 
\begin{enumerate}
\item Quantum Annealing can be used to solve the feature selection task and the quality of the solutions obtained for both classification and ranking problems, either in terms of accuracy or NDCG, is comparable with the quality obtained with traditional solvers;
\item for problems able to fit within the number of available qubits, Quantum Annealing requires less time than any other traditional solver, while for very large problems the Hybrid quantum-classical solver is faster than traditional solvers.
\end{enumerate}

From a broader perspective, this study provides evidence that Quantum Annealer technology has evolved to the point that it can be used to tackle real problems. 
This, in combination with the simple computing model and the easy access to Quantum Annealers in the cloud opens new research directions in the application of this technology to tackle computationally intensive tasks in many \ac{IR} domains.
There is a significant need to identify which are the tasks that fit well in a \ac{QUBO} formulation or can be efficiently approximated into one by leveraging or developing new heuristics. 
Many open questions also lie in understanding how to efficiently perform the minor embedding phase, especially for problems that do not have the regular structure of the fully-connected feature selection one. As the technology and tools improve, it is easy to imagine a library of precomputed embeddings being available for problems of particular structures in a similar way as how pretrained machine learning models are. This would completely remove the computational cost of generating the embedding for those cases.
Another important direction of improvement is to reduce the effects of network latency, which will be minimized when quantum technology is integrated into the low-latency networks of data centers.
Finally, the impact of advanced Quantum Annealer controls and advanced annealing schedule (\eg reverse annealing \cite{DBLP:journals/corr/abs-2007-05565}), that really bring the researcher or practitioner close to the physics of the underlying system, can have a strong impact on the solution quality or on the likelihood of finding a good solution but are not yet well understood.  

Overall, this work has shown that \acf{QUBO} and \acf{QA} are viable options for improving feature selections for both classification and ranking and the above discussion on future perspectives gives an idea of how much room for improvement is already possible to imagine. Therefore, it would be definitely worth if we, as a community, undertake a systematic exploration of these promising research directions, not forgetting that while feature selection is a specific task, for other relevant tasks as well it may be possible to develop a formulation suitable for applying quantum computing approaches.
Ranking and classification are central not only to IR but also to several neighbourhood areas, such as natural language processing and recommender systems. Therefore, we could promote some joint effort across these communities, in order to maximize the impact and benefit from cross-fertilization. In this respect, \ac{IR} has an extremely long tradition in community-wide cooperation on shared research activities, very successfully embodied by large scale evaluation campaigns, as TREC, CLEF, NTCIR and FIRE. It would be extremely valuable if such initiatives take a lead and promote the organization of shared activities for exploring the application of quantum computing to IR, NLP, and RecSys in a comparable and shared way.

\section{Acknowledgments}
We acknowledge the CINECA award under the ISCRA initiative, for the availability of quantum computing resources and support. The work was partially supported by University of Padova Strategic Research Infrastructure Grant 2017: “CAPRI: Calcolo ad Alte Prestazioni per la Ricerca e l’Innovazione”.


\bibliographystyle{ACM-Reference-Format}
\balance
\bibliography{main-references}

\end{document}